\pdfoutput=1
\documentclass[reprint,amsmath,amssymb,aps]{revtex4-1} 

\usepackage{graphicx}
\usepackage{dcolumn}
\usepackage{bm}
\usepackage[bookmarks=false]{hyperref}
\usepackage{color}
\usepackage{ulem}

\begin{document}


\title{Hybrid waves localized at hyperbolic metasurfaces}



\author{O. Yermakov}
\altaffiliation{V.N. Karazin Kharkiv National University, Kharkiv 61022, Ukraine}
\author{A. Ovcharenko}
\altaffiliation{V.N. Karazin Kharkiv National University, Kharkiv 61022, Ukraine}
\author{A. Bogdanov}
\email{bogdanov@mail.ioffe.ru}
\altaffiliation{Ioffe Institute, St.~Petersburg 194021, Russia}
\altaffiliation{Academic University, St.~Petersburg 194021, Russia}
\author{I. Iorsh}
\email{i.iorsh@phoi.ifmo.ru}
\author{Yu.~S. Kivshar}
\altaffiliation{Nonlinear Physics Center, Australian National University, Canberra ACT 0200, Australia}
\affiliation{ITMO University, St.~Petersburg 197101, Russia}
\date{\today}

\begin{abstract}
We reveal the existence of a new type of surface electromagnetic waves supported by hyperbolic metasurfaces, described
by a conductivity tensor with an indefinite signature. We demonstrate that the spectrum of the hyperbolic metasurface waves
consists of two branches corresponding to hybrid TE-TM waves with the polarization that varies from linear to elliptic or circular depending on the wave frequency and propagation direction.  We analyze the effect of losses of the surface waves
and derive the corresponding analytical asymptotic expressions.
\end{abstract}

\pacs{Valid PACS appear here}
\maketitle


\section{\label{sec:introduction}Introduction}

\textit{Metasurfaces} are known as a two-dimensional  analogue of metamaterials, and they offer unprecedented control over the light propagation, reflection, and refraction~\cite{yu2014flat, kildishev2013planar}. One of the main advantages of metasurfaces is that these structures are fully compatible with the modern planar fabrication technology and they can be readily integrated into the on-chip optical devices preserving the most of functionalities  of three-dimensional metamaterials. In a general case, a metasurface can be described as a two-dimensional current characterized by a dispersive nonlocal two-dimensional conductivity tensor. At the same time, electromagnetic properties of a broad and constantly growing class of two-dimensional materials such as graphene, silicene, hexagonal boron nitride, can also be characterized by conductivity tensors. Thus, the physics of metasurfaces and optics of two-dimensional materials are tightly interconnected. Particularly, it has been shown~\cite{mikhailov2007new} that a graphene sheet can support transverse electric (TE) surface polaritons in the frequency region where the imaginary part of the surface conductivity becomes negative. Negative imaginary part of the conductivity corresponds to the capacitive surface impedance. At the same time, the existence of the TE surface waves at the capacitive impedance surfaces has been studied previously~\cite{sievenpiper1999high}.

In this paper, we study a special class of metasurfaces characterized by a local diagonal anisotropic conductivity tensor. Such metasurfaces can be regarded as a two-dimensional analogue of uniaxial crystals. Specifically, when the imaginary parts of the principal components of the conductivity tensors have different signs, a strong correspondence appears between these structures and hyperbolic metamaterials~\cite{poddubny2013hyperbolic}. Here we focus on the dispersion and polarization properties of the localized waves supported by these metasurfaces.

It was shown that an anisotropic interface separating a hyperbolic metamaterial and vacuum can support a certain class of plasmonic modes analogous to the Dyakonov surface states~\cite{artigas2005dyakonov, jacob2008optical,Takayama2008,Polo2011}. Dyakonov surface states~\cite{Dyakonov1988} are localized modes which can propagate in a narrow angle range along the interface of the anisotropic crystals. Despite the theoretical prediction back in the 1980s, these modes have been experimentally demonstrated only recently~\cite{takayama2009observation}. This is due to the fact that  for the case of an interface of conventional anisotropic crystal these modes can propagate only in an extremely narrow range of angles, and thus it is hard to excite them experimentally. Nevertheless, these modes attract a significant scientific interest since they suggest a route for  virtually lossless optical information transfer at the nanoscale, which is extremely important for the perspectives of on-chip optical data processing devices. Moreover, as was shown recently, the propagation direction of these modes can be effectively controlled by slight modification of the dielectric permittivities of the structure~\cite{takayama2014lossless}.

Here we show that, in a sharp contrast to bulk hyperbolic metamaterials, the hyperbolic metasurfaces can support two types of surface modes at a single frequency. These modes originate from the coupling of the TM and TE polarized surface polaritons. A similar effect occurs in graphene sheets in the presence of a strong magnetic field perpendicular to the graphene layer~\cite{Iorsh2013} or on a metal substrate coated by a thin anisotropic dielectric film~\cite{abdulhalim2009surface}. The corresponding surface waves have elliptic polarization which is essential for the construction of on-chip optical networks~\cite{karkar2013hybrid,karkar2012surface,ohashi2009chip}.

\section{\label{sec:model}Model}

\subsection{Dispersion equation} 

\begin{figure*}[htbp]
   \centering
   \includegraphics[scale=0.33]{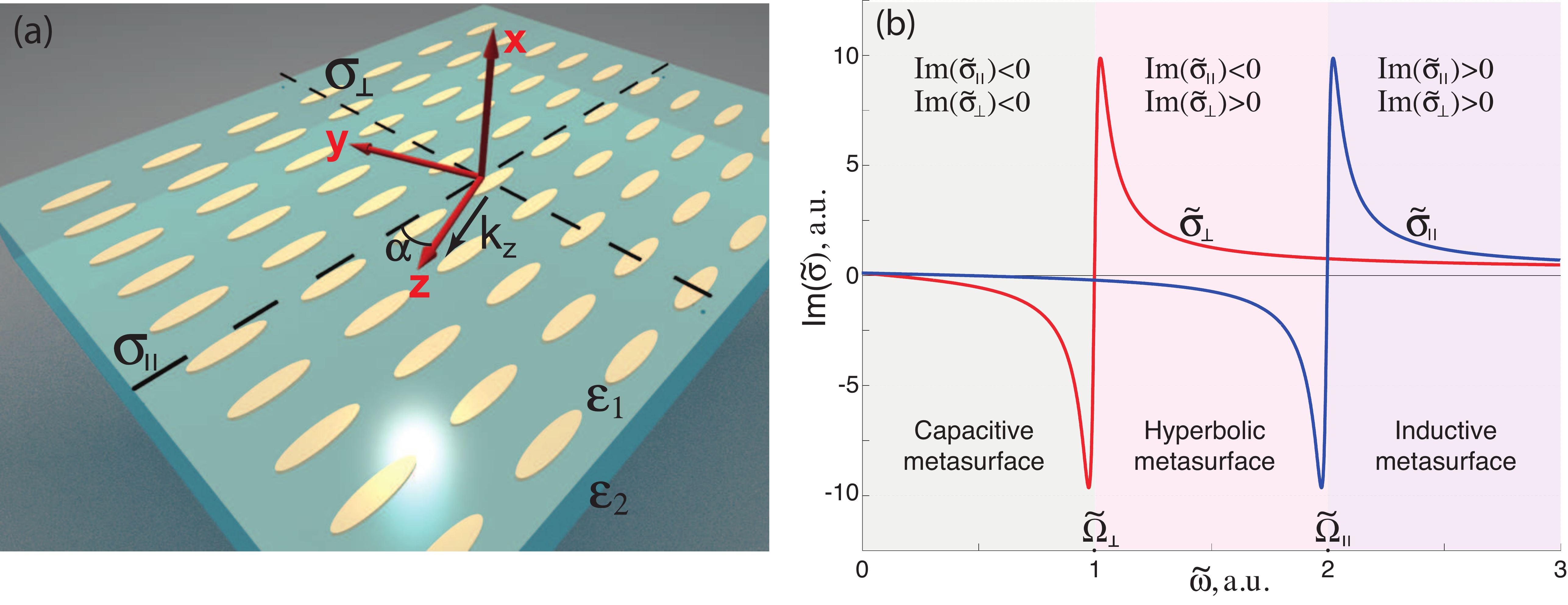} 
   \caption{(a) Geometry of the problem. Upper ($x>0$) half space with $\varepsilon_1$ and lower ($x<0$) half space with $\varepsilon_2$ are separated by an anisotropic conducting layer. The principal axes of the conductivity tensor are shown by dashed lines. Surface wave propagates along $z$-direction. (b) Frequency dependence  of imaginary parts of dimensionless conductivity tensor components $\widetilde{\sigma}_\bot$ and $\widetilde{\sigma}_\|$. Parameters of conductivity tensor components are following: $\widetilde{\Omega}_\bot=1$, $\widetilde{\Omega}_\|=3$, $\widetilde{\gamma}=0.05$.}
   \label{Sketch}
\end{figure*}
We consider two isotropic media with permittivities $\varepsilon_1$ and $\varepsilon_2$ separated by an anisotropic non-chiral metasurface [Fig.   \ref{Sketch}(a)]. {In the figure we suggest one of the possible design of hyperbolic metasurface representing a two-dimensional array of asymmetric subwavelength resonators. Within the local homogenization approach the electromagnetic properties of such structure can be described by a two dimensional conductivity tensor:}

\begin{equation}
\widehat\sigma_0=\left(
\begin{matrix}
\sigma_\bot & 0 \\
0 &\sigma_\|
\end{matrix}
\right).
\label{sigma_zero}
\end{equation}
Here $\sigma_\bot$ and $\sigma_\|$ are frequency dependent conductivities per unit length corresponding to the principal axes of the tensor.

We will seek solution of the Maxwell's equations in the form of a traveling wave propagating in the $z$-direction and localized in the $x$-direction. The both electric and magnetic fields depend on the $z$-coordinate and time $t$ as $\exp(ik_zz-i\omega t)$. We assume that $\alpha$ is the angle between $z$-direction and one of principle {axes} of the tensor [see Fig. \ref{Sketch}(a)]. Conductivity tensor is not diagonal in the chosen set of coordinates and can be written as
\begin{eqnarray}
\hat{\sigma} = \begin{pmatrix}
 \sigma_{yy} & \sigma_{yz}\\
 \sigma_{zy} & \sigma_{zz}
\end{pmatrix},
\end{eqnarray}
where
\begin{eqnarray}
\sigma_{yy}&=&\sigma_{\bot}\cos^2{\alpha} + \sigma_{\|}\sin^2{\alpha}; \\
\sigma_{zz}&=&\sigma_{\bot}\sin^2{\alpha} + \sigma_{\|}\cos^2{\alpha}; \\
\sigma_{yz}&=&\sigma_{zy}=\dfrac{\sigma_{\|} - \sigma_{\bot}}{2}\sin{2\alpha}.
\end{eqnarray}

Electric and magnetic fields  (${\bf E}$ and ${\bf H}$) obey the following boundary conditions on the metasurface:
\begin{eqnarray}
\label{H_bc}
{\bf [n,H}_2] - {\bf[n,H}_1] &=& \frac{4\pi}{c}\hat{\sigma}{\bf E};\\
\label{E_bc}
{\bf [n,E}_2] - {\bf[n,E}_1] &=& 0.
\end{eqnarray}
{Index} $1$ ($2$) corresponds to the upper (lower) half-space, ${\bf n}$ is a unit vector normal to the interface.

Dispersion equation for surface waves can be obtained from Maxwell's equations using boundary conditions (\ref{H_bc}) and (\ref{E_bc}) and condition that electromagnetic field decays away from the interface
\begin{align}
\label{disp_eq}
&\left( \frac{c\kappa_1}{\omega} + \frac{c\kappa_2}{\omega} - \frac{4\pi i}{c}\sigma_{yy}\right) \nonumber \\ \times & \left( \frac{\omega\varepsilon_1}{c\kappa_1} + \frac{\omega\varepsilon_2}{c\kappa_2} + \frac{4\pi i}{c}\sigma_{zz} \right) = \frac{16\pi^2}{c^2}\sigma_{yz}^2.
\end{align}
Here $\kappa_{1,2}^2=k_z^2-\varepsilon_{1,2}\omega^2/c^2$ is inversed penetration depth of the surface wave into upper and lower medium. {Similar equation describes dispersion of magnetoplasmons,  surface waves in a two-dimensional electron gas  in the presence of a strong DC magnetic field \cite{ferreira2012confined, chiu1974plasma, Iorsh2013}.}

\subsection{Conductivity tensor}

 Before analyzing dispersion equation (\ref{disp_eq}) let us discuss the form of conductivity tensor components $\sigma_{\bot}$ and $\sigma_{\|}$. Conductivity tensor describes a response of the layer on external electromagnetic field. In the common case, the response is not local in space and time that results in dependence of the conductivity on the frequency $\omega$ and wave vector $k_z$. Equation (\ref{disp_eq}) was obtained using spatially local boundary conditions (\ref{H_bc}, \ref{E_bc}). Therefore, direct substitution of $\sigma_\bot$ and $\sigma_\|$ {as function of} $k_z$ into Eq.~ (\ref{disp_eq}) is not correct whereas Eq.~ (\ref{disp_eq}) is still correct for frequency dependent $\sigma_\bot$ and $\sigma_\|$.

Let us put forward the assumption that conductivity tensor components depend on the frequency according to the Drude-Lorentz model:
\begin{equation}
\sigma_s(\omega)=A\frac{ic}{4\pi}\frac{\omega}{\omega^2-\Omega_s^2+i\gamma\omega}, \ \ s=\bot,\|.
\label{sigma_disp}
\end{equation}
Such dispersion shape is quite natural to many systems in optical, infrared, THz, and radio frequency ranges \cite{Jackson1962}. In Eq.~(\ref{sigma_disp}) $\Omega_s$ are resonance frequencies along the principle axes of the conductivity tensor, $\gamma$ is a relaxation constant which is supposed to be isotropic. Constant $A$ has a dimension of $rad/s$. Explicit expression of A is defined by  the metasurface design. This constant can be excluded from the analysis with the  help of the following dimensionless variables:
\begin{equation}
\begin{split}
\widetilde{\sigma}_s=\frac{4\pi\sigma_s}{c}; \ \widetilde{\omega}=\frac{\omega}{A}; \ \widetilde{\gamma}=\frac{\gamma}{A}; \\
\widetilde{\kappa}=\frac{c\kappa}{A\sqrt{\varepsilon}}; \ \widetilde{k}_z=\frac{ck_z}{A\sqrt{\varepsilon}}. \ \ \
\end{split}
 \end{equation}
Real part of $\widetilde{\sigma}_\bot$ and $\widetilde{\sigma}_\|$ is responsible for energy dissipation and imaginary part for the polarizability of the structure. Typical frequency dependence of $\text{Im}(\widetilde{\sigma}_\bot)$ and $\text{Im}(\widetilde{\sigma}_\|)$ is shown in {Fig.  \ref{Sketch}(b)}. One can see that signature of conductivity tensor (\ref{sigma_zero}) depends on the frequency. {It is possible to distinguish three cases: (i) capacitive metasurface when both $\text{Im}(\widetilde{\sigma}_{\bot})$ and $\text{Im}(\widetilde{\sigma}_{\|})$ are negative; (ii) hyperbolic metasurface when $\text{Im}(\widetilde{\sigma}_\bot)\text{Im}(\widetilde{\sigma}_\|)<0$; (iii) inductive metasurface when both $\text{Im}(\widetilde{\sigma}_{\bot})$ and $\text{Im}(\widetilde{\sigma}_{\|})$ are positive}.


\section{Results and discussion \label{res_and_discuss}}

\subsection{Dispersion of surface waves} 
\begin{figure}[htbp]
   \centering
   \includegraphics[scale=0.45]{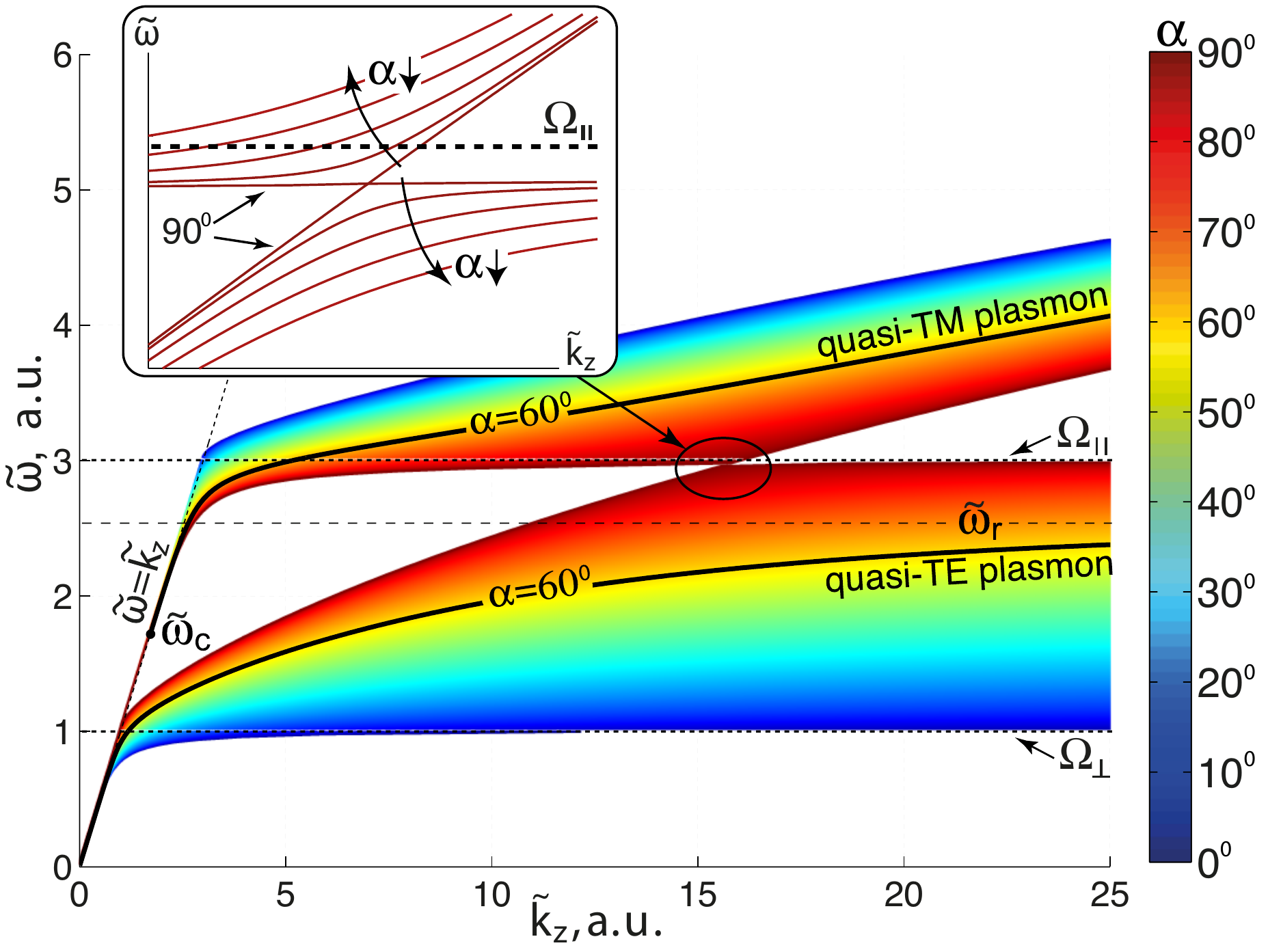} 
   \caption{Dependence of $\widetilde{k}_z$ on $\widetilde{\omega}$ for the surface waves on hyperbolic metasurface for different propagation directions $\alpha$. Two branches correspond to quasi-TM and quasi-TE surface plasmons. The inset shows the structure of dispersion curves at $\alpha\approx90^0$.}
   \label{rainbow}
\end{figure}

In order to analyze dispersion of surface waves which is described by Eq.~ (\ref{disp_eq}) let us neglect dissipation of the energy in the system and put $\widetilde{\gamma}=0$. The case $\gamma\neq0$ is analyzed in Sec.~\ref{sec:losses}. For the sake of simplicity, further we will consider symmetric situation when $\varepsilon_1=\varepsilon_2=\varepsilon$. As an example, let us consider the structure  with resonance frequencies of conductivity tensor components $\widetilde{\Omega}_\bot=1$ and $\widetilde{\Omega}_\|=3$. Dependence of wave vector $k_z$ on frequency $\omega$ and propagation direction $\alpha$ can be obtained analytically from Eq.~(\ref{disp_eq}). This equation yields two solutions which correspond to hybrid polarized waves (quasi-TE and quasi-TM plasmons). Their dispersion for $\alpha=60^o$ is shown in Fig.  \ref{rainbow}.


Surface waves of pure TE or TM polarization can propagate only along the principle axes of the conductivity tensor ($\alpha=0^o$ and $\alpha=90^o$). In this case, right part of Eq.~ (\ref{disp_eq}) is equal to zero and the equation splits into two independent equations corresponding to two-dimensional TE and TM plasmons \cite{luo2013plasmons}.

Frequency cutoff of quasi-TE plasmon is equal to zero and does not depend on $\alpha$.  Frequency cutoff $\omega_{c}$ of quasi-TM plasmon belongs to the interval $\Omega_\bot\leqslant\omega_c\leqslant\Omega_\|$ and depends on $\alpha$. This dependence can be found from the equation:
\begin{equation}
\cot^2 \alpha =-\frac{\sigma_\|(\omega_{c})}{\sigma_\bot(\omega_{c})}.
\label{cutoff}
\end{equation}
Quasi TE-plasmon has a maximal frequency $\omega_{r}$ at which it can propagate. Dependence of $\omega_{r}$ on the $\alpha$ yields by the equation:
\begin{equation}
\tan^2 \alpha =-\frac{\sigma_\|(\omega_{r})}{\sigma_\bot(\omega_{r})}.
\label{om_max}
\end{equation}
It follows from Eqs.~(\ref{cutoff}) and (\ref{om_max}) that simultaneous propagation of both surface waves at the same frequency is possible only if
\begin{equation}
\frac{\pi}{4}\leqslant|\alpha|\leqslant\frac{3\pi}{4}.
\label{necess_cond}
\end{equation}
This condition does not depend on the specific dispersion of $\widetilde{\sigma}_{\bot,\|}$ and can be fulfiled for any hyperbolic metasurface {\footnote{It is necessary to mention that condition (\ref{necess_cond}) is necessary but not sufficient. For the sufficiency, Eqs.~ (\ref{cutoff}) and (\ref{om_max}) must be solvable for $\alpha$ from Eq.~ (\ref{necess_cond}).}}.

 \begin{figure*}[htbp]
   \centering
  \includegraphics[scale=0.9]{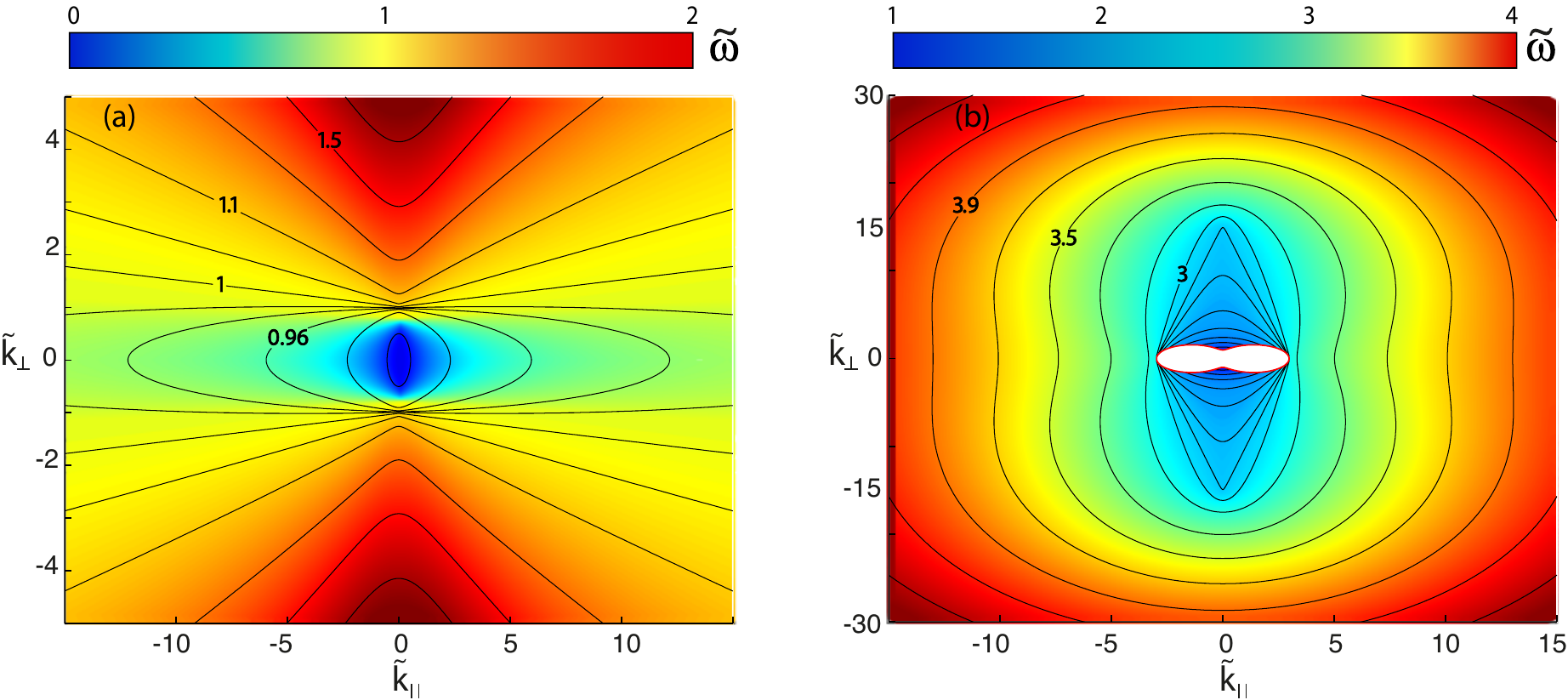} 
   \caption{Contours of equal frequency on $\mathbf{k}$-plane for quasi-TE (a) and quasi-TM (b) surface plasmons. Here $\widetilde{k}_\bot$ and $\widetilde{k}_\|$ are component of the wave vector along the principle axes of the conductivity tensor.}
   \label{iso1}
\end{figure*}

Dispersion curves of quasi-TM and TE  plasmons for all angles are shown in Fig.  \ref{rainbow}. It follows from Eq.~  (\ref{disp_eq}) and can be seen from the figure that there is a frequency gap between $\Omega_\bot$ to $\Omega_\|$ at the angle $\alpha=0^o$. The gap is squeezed as $\alpha$ tends to $\pi/2$. Structure of the dispersion curves at $\alpha$ {close} to $\pi/2$ is shown in the inset of Fig.  \ref{rainbow}. Presence of the anti-crossing means that there is a polarization switching along the dispersion curves. Therefore, notation "quasi-TM" and "quasi-TE" plasmons is just formal.

Figure \ref{rainbow} contains full information about the dispersion of the surface waves on hyperbolic metasurface, but some peculiarities of the wave propagation associated, for example, with density of optical states or relative direction of phase and group velocities are best understood in ${\mathbf k}$-space using isofrequency contours. Isofrequency contours for quasi-TE and quasi-TM plasmons are shown in Fig.  \ref{iso1}. One can see that for quasi-TE plasmons, the contours have elliptical, $\infty$-shape or hyperbolic form depending on the frequency. Contours of equal frequency for quasi-TM plasmons have arc, rhombus, 8-shaped or elliptical form depending on the frequency. The arc contours is observed for the hyperbolic regime when $\widetilde{\Omega}_\bot<\widetilde{\omega}<\widetilde{\Omega}_\|$. End points of the arc correspond to the frequency cutoffs $\widetilde{\omega}_c$ which obey Eq.~  (\ref{cutoff}). Its solution represents fourth-order curve in $\mathbf{k}$-plane, so-called \textit{hippopede} \cite{Lawrence2013}:
\begin{equation}
\widetilde{\omega}^2_c=\widetilde{\Omega}_\bot^2\sin(\alpha)^2+\widetilde{\Omega}_\|^2\cos(\alpha)^2.
\label{hipopede}
\end{equation}
Discontinuity of the equal frequency contours at the finite points which takes place for the arcs in our case is unusual for bulk waves in three dimensional space but can be observed for the surface ones \cite{Dyakonov1988}. Inside the hipopede quasi-TM plasmons are leaky-modes.

\subsection{Effect of losses \label{sec:losses}} 

\begin{figure}[htbp]
   \centering
   \includegraphics[scale=0.5]{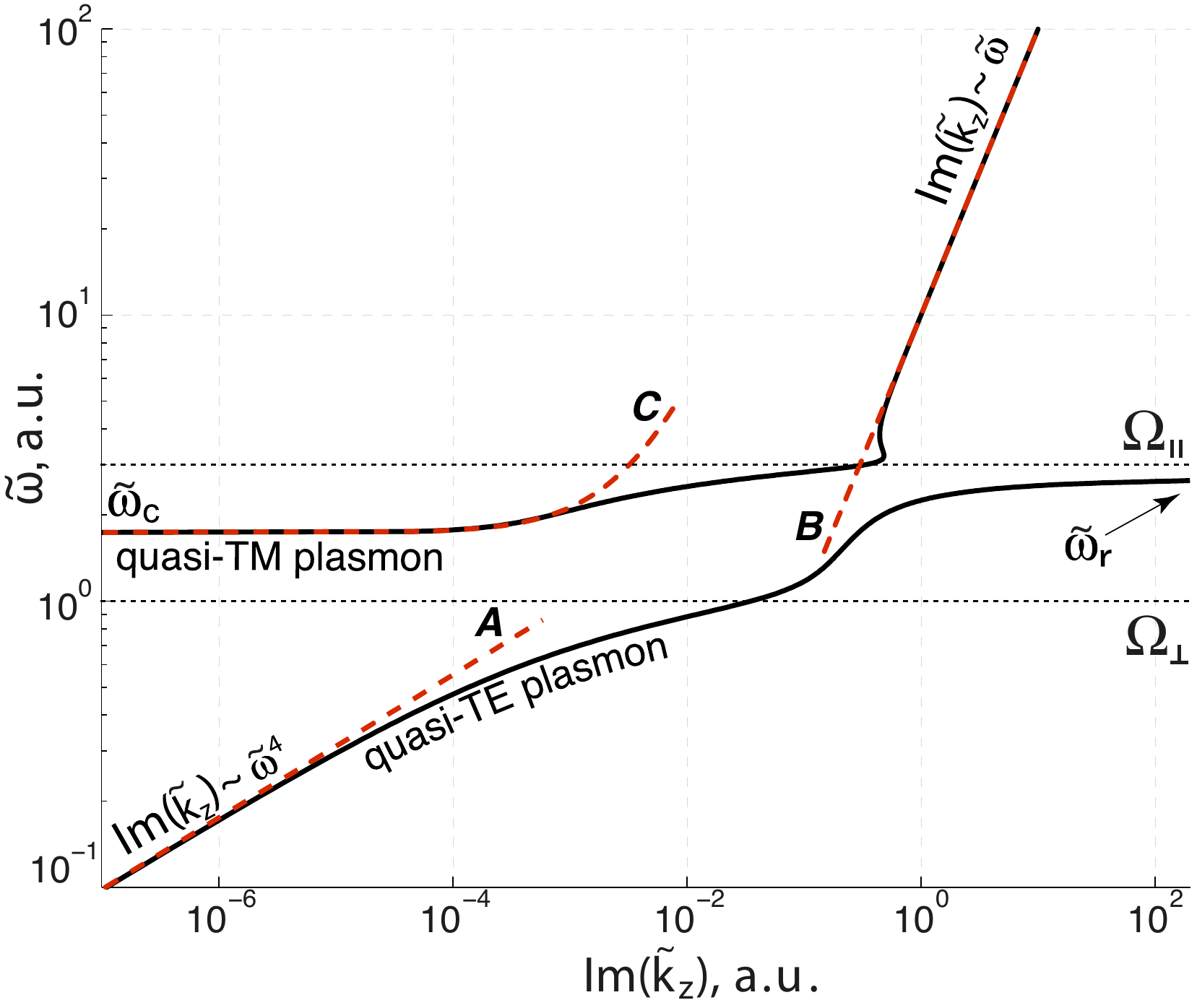} 
   \caption{Dependence of $\text{Im}(\widetilde{k}_z)$ on $\widetilde{\omega}$ for quasi-TE and quasi-TM surface plasmons for $\alpha=60^0$ plotted on log-log scale. Dashed lines are asimptotics of the losses described. Curve \textbf{\textit{A}} is described by Eq.~ (\ref{lowfreq_asympt}); curve \textbf{\textit{B}} is described by Eq.~(\ref{highfreq_asympt}); curve \textbf{\textit{C}} is described by Eq.~(\ref{asymp_cutoff}).}
   \label{fig_losses}
\end{figure}

Let us take into account energy dissipation and put $\widetilde{\gamma}\neq0$  in Eq.~ (\ref{sigma_disp}).  It results in finite propagation length of the surface waves which is proportional to $1/\text{Im}(\widetilde{k}_z)$. Frequency dependence of $\text{Im}(\widetilde{k}_z)$ for quasi-TE and TM plasmons is shown in Fig.  \ref{fig_losses}.  Using log-log scale makes it obvious that $\text{Im}(\widetilde{k}_z)\sim\widetilde{\omega}^4$ at $\widetilde{\omega}\ll\widetilde{\Omega}_\bot$ and $\text{Im}(\widetilde{k}_z)\sim\widetilde{\omega}$ at $\widetilde{\omega}\gg\widetilde{\Omega}_\|$. Dependence of $\text{Im}(\widetilde{k}_z)$ on $\widetilde{\omega}$  can be obtained analytically from (\ref{disp_eq}) using perturbation theory. In the case of $\widetilde{\omega}\ll\widetilde{\Omega}_\bot$ the losses can {be} written as
\begin{equation}
\text{Im}(\widetilde{k}_z)\!=\!\! \frac{\widetilde{\gamma}\widetilde{\omega}^4}{4}\!\left(\frac{\cos^2\!\alpha}{\Omega^2_\bot}\!+\!\frac{\sin^2\!\alpha}{\Omega^2_\|}\right)\!\!\left(\frac{\cos^2\!\alpha}{\Omega^4_\bot}\!+\!\frac{\sin^2\!\alpha}{\Omega^4_\|}\right).
\label{lowfreq_asympt}
\end{equation}
In the case of $\widetilde{\omega}\gg\widetilde{\Omega}_\|$ the losses can be written as
\begin{equation}
\text{Im}(\widetilde{k}_z)= 2\widetilde{\gamma}\widetilde{\omega}.
\label{highfreq_asympt}
\end{equation}

One can see that in contrast to the case of low frequencies, the losses do not depend on the propagation direction $\alpha$ and resonances frequencies $\widetilde{\Omega}_\bot$ and $\widetilde{\Omega}_\|$.

In the vicinity of the frequency cutoff $\widetilde{\omega}_c$ frequency dependence of $\text{Im}(\widetilde{k}_z)$ can be represented as
\begin{equation}
\text{Im}(\widetilde{k}_z)=\frac{\widetilde{\gamma}\widetilde{\omega}_c}{2}\frac{\widetilde{\sigma}_\|^2(\widetilde{\omega}_c)\sin^2(\alpha)+\widetilde{\sigma}_\bot^2(\widetilde{\omega}_c)\cos^2(\alpha)}{1+\widetilde{\sigma}_\|(\widetilde{\omega}_c)\widetilde{\sigma}_\bot(\widetilde{\omega}_c)/4}\delta\widetilde{\omega}.
\label{asymp_cutoff}
\end{equation}
\begin{figure}[htbp]
   \centering
   \includegraphics[scale=1.2]{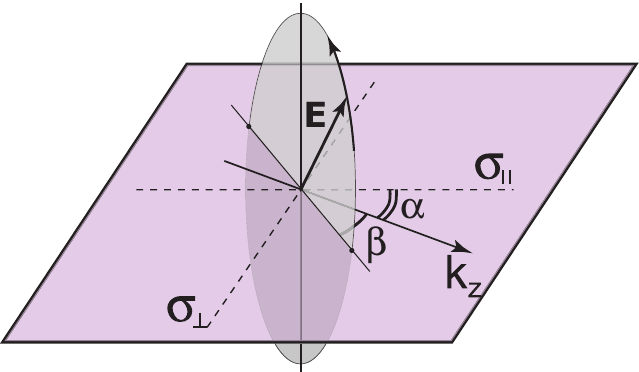} 
   \caption{Polarization structure of hybrid surface waves on the hyperbolic metasurface.}
   \label{fig_polarization}
\end{figure}
Here $\delta\widetilde{\omega}=\widetilde{\omega}-\widetilde{\omega}_c$ and $\delta\widetilde{\omega}\ll\widetilde{\omega}_c$. One can see from Eq.~(\ref{asymp_cutoff}) that $\text{Im}(\widetilde{k}_z)\rightarrow0$ as $\delta\widetilde{\omega}\rightarrow0$. It is a result of weak localization of quasi-TM plasmon near the frequency cutoff $\widetilde{\omega}_c$. The opposite situation occurs for quasi-TE plasmon mode near $\widetilde{\omega}_r$ where this mode is strongly localized.


\subsection{Polarization properties} Spatial inhomogeneity and partly longitudinal polarization of surface waves results in {an} unusual spatial distribution of their momentum and angular momentum density. In particular, surface waves can {possess} transversal Belinfante's spin momentum \cite{belinfante1940current, Bliokh2014}. Anisotropy of the hyperbolic metasurface results in mixing of TE and TM surface waves and makes their polarization structure very manifold.

It can be shown from the Maxwell's equations that for the surface waves under consideration electric field components $E_z$ and $E_y$ are in-phase and $E_x$ has $\pi/2$ phase delay. It means that electric field $\mathbf{E}$ rotates in the plane orthogonal to the metasurface so that the end of $\mathbf{E}$  draws an ellipse. In the common case, the plane of the ellipse is rotated on the angle $\beta$ with respect to the wave vector $\widetilde{k}_z$ (Fig.  \ref{fig_polarization}).

At low frequencies ($\widetilde{\omega}\ll\widetilde{\Omega}_\bot$), there is only quasi-TE mode. Therefore, angle $\beta$ is close to 90$^0$ and the ellipse is almost degenerated into a line segment. At high frequencies, where only quasi-TM mode propagates, angle $\beta$ is near 0$^0$ and the ellipse represents a circle.

Strong hybridization of quasi-TE and quasi-TM modes plasmon results in their unusual polarization. For example, for quasi-TM plasmon at $\alpha=11.5^0$ and $\widetilde{\omega}=2.95$ angle $\beta=87.3^0$ and difference between semi-axis of the ellipse {is} less than 4\%. Therefore, the wave has circular polarization. Absorption for this wave is quite low due to the vicinity {of} frequency cutoff $\widetilde{\omega}_c$. In this case,  figure of merit (FOM) can be estimated using Eq.~(\ref{asymp_cutoff}) or numerically

\begin{equation*}
\text{FOM}=\frac{\text{Re}(\widetilde{k}_z)}{\text{Im}(\widetilde{k}_z)}\approx1\cdot10^5 \ \  \text{for} \ \ \widetilde{\gamma}=0.05.
\end{equation*}

\section{\label{sec:conclusion}Concluding remarks}

We have presented a comprehensive analysis of surface waves propagating along hyperbolic metasurfaces. We have analysed dispersion, losses, and polarization properties of such waves in the most general form, not specifying a specific design of the metasurface and describing its properties using the effective conductivity approach. Within this approach, the problem does not acquire a specific scale and, therefore, the results can be applied to different frequencies ranging from microwaves to ultraviolet band. 

We have shown that the spectrum of waves supported by hyperbolic metasurfaces consists of two branches of hybrid TE-TM polarized modes, that can be classified as quasi-TE and quasi-TM plasmons. Dispersion properties of these waves are strongly anisotropic, and they have some similar features with magnetoplasmons and two-dimensional TE and TM plasmons.

Analytical solutions of the problem allow detailed study of the surface wave properties. So, simple asymptotic equations for the losses have been obtained near the frequency cutoff of quasi-TM plasmon mode and in the high and low frequency regions. Analysis of the equal frequency contours shows that their form and topology drastically depend on the frequency. The contours can have elliptic, hyperbolic, 8-shaped, rhombus, or arc form. Multiplicity of equal frequency contours allows to forecast in hyperbolic metasurface such phenomena as negative refraction  \cite{Luo2002}, self-collimation \cite{Kosaka1999, Witzens2002, Stein2012}, channeling  of surface waves \cite{Chigrin2003, Belov2005},  and large  spontaneous emission enhancement of the quantum emitters due to the large density of states \cite{poddubny2012microscopic, Sreekanth2013, Sreekanth2014}.

We have shown that hyperbolic metasurfaces support simultaneous propagation of both quasi-TE and quasi-TM plasmon surface modes at the same frequency, and we derive the specific conditions for this to occur. Neither in isotropic nor in chiral metasurfaces such a phenomenon is known \cite{Iorsh2013}. Polarization structure of the surface waves can vary substantially so the polarization can change from linear to circular with different orientation of polarization plane.

Unique electromagnetic properties of hyperbolic metasurfaces make them quite promising for applications in many areas such as resonance sensing and detection, superlensing and near-field imaging, enhanced Raman spectroscopy, optical antennas, on-chip optical networks, etc. Taking into account their fabrication simplicity, rich functionality, and planar geometry it is possible to assert that hyperbolic metasurfaces can be a basis of many optical and optoelectronic devices.

\acknowledgments

This work was partially supported by the Government of the Russian Federation (Grant 074-U01), the Australian Research Council, and the program on Fundamental Research in Nanotechnology and Nanomaterials of the Presidium of the Russian Academy of Sciences. A.B. appreciates the valuable support of RFBR (grant No.~14-02-01223). O.Y. acknowledges the Dynasty Foundation. 

\bibliography{References_bogdanov}

\end{document}